\documentclass[prl,superscriptaddress,twocolumn]{revtex4}
\usepackage[dvips]{graphicx} 
\usepackage{latexsym}
\begin{document}

\newcommand{\fig}[2]{\includegraphics[width=#1]{#2}}

\title{Field theory of statics and dynamics of glasses: rare events and barrier
distributions
}
\author{Leon Balents} 
\affiliation{Department of Physics, University of California, Santa Barbara, CA 93106--4030} 
\author{Pierre Le Doussal} 
\affiliation{CNRS-Laboratoire de Physique Th\'eorique de l'Ecole Normale Sup\'erieure,
24 Rue Lhomond 75231 Paris,
France} 

\date{\today}

\begin{abstract}
  We study thermally activated dynamics using functional
  renormalization within the field theory of randomly pinned elastic
  systems, a prototype for glasses.  It appears through an essentially
  non-perturbative boundary layer in the running effective action, for
  which we find a consistent scaling ansatz to all orders.  We find
  that in the statics the boundary layer describes the physics of rare
  low energy metastable states as suggested by a phenomenological
  droplet picture.  The extension to equilibrium dynamics mandates a
  non-trivial broad distribution of relaxation times for the large
  scale modes, hitherto neglected in analytic calculations.
\end{abstract}

\maketitle

\vspace{0.15cm}


Extremely slow dynamics and dominance of rare fluctuations are
ubiquitous properties of complex and disordered materials.  Such
behavior occurs in a panoply of systems involving many
interacting degrees of freedom, from spin glasses to supercooled
liquids to the pinned elastic medium.  We study the latter here as
a prototype due to its relative simplicity.  It describes
numerous experimental systems such as interfaces in magnets
\cite{review,creepexp}, charge density waves and Wigner crystals \cite{gruner_revue_cdw},
or vortex lattices in
superconductors \cite{vortices}.  The elastic medium
is described by a displacement field $\Phi({\bf r})$, with Hamiltonian
\begin{equation}
  H[u] = \int \! d^d{\bf r} \left[ {1 \over 2}|\nabla \Phi|^2 +
    V(\Phi({\bf r}),{\bf r}) \right], \label{eq:ham}
\end{equation}
where $V(\Phi,r)$ is a zero-mean
Gaussian random potential with variance
$\overline{V(\Phi,r)V(\Phi',r')} = R(\Phi-\Phi')\delta^{(d)}(r-r')$
(an overline indicates disorder averaging).  Different physical
cases are described by varying longitudinal
dimensionality ($d$), transverse dimensionality ($N=$ the number of
vector components of $\Phi$ -- we take $N=1$ for simplicity here), and
$R(\Phi)$.  The Langevin equation of
motion is
\begin{eqnarray}
  \eta \partial_t \Phi_{rt} = \nabla^2_r \Phi_{rt} + f(\Phi_{rt},r) +
  \zeta(r,t), \label{eom}
\end{eqnarray}
with friction $\eta$, thermal noise
$\zeta(r,t)$
and random force $f(\Phi,r) = - \partial V(\Phi,r)/\partial \Phi$
(denoting $\Phi(r,t)\equiv\Phi_{rt}$).  

A minimal physical description of disordered glasses is provided by
the droplet scenario \cite{droplets}. In its simplest form, it
supposes the existence, at each length scale $L$, of a small number of
excitations of size $\delta\Phi\sim L^\zeta$ above a ground state,
drawn from an energy distribution of width $\delta E \sim L^\theta$
with constant weight near $\delta E=0$.  Although the applicability of
the droplet picture to more complex systems such as spin glasses is
controversial, it appears to describe simpler models such as
Eq.~(\ref{eq:ham}) relatively well, at least in low
dimensions,\cite{dropevid}\ and we summarize its main conclusions.
Static thermal fluctuations at a given scale are dominated by rare
samples/regions with two nearly degenerate minima.  In particular, the
$(2n)^{\rm th}$ moment of $\Phi$ fluctuations is expected to behave as
\begin{equation}
  \overline{(\langle \Phi^2\rangle - \langle \Phi\rangle^2)^n} \sim
  c_n (T/L^\theta) L^{2n\zeta}. \label{eq:twowell}
\end{equation}
The droplet picture supposes the long-time equilibrium dynamics is
dominated by thermal activation between these quasi-degenerate minima
controlled by barriers of typical scale $U_L\sim L^\psi$.  Little is
known about the distribution of these barriers, but there is some
evidence \cite{drossel_barrier} that $\psi\approx \theta$.  Even a modest
distribution of barriers, however, due to the Arrhenius law $\tau_L \sim
e^{U_L/T}$, yields relaxation time scales with an extremely broad
distribution as $T\rightarrow 0$.  An important unanswered question is
the behavior of relaxation time moments, defined e.g. by
\begin{equation}
  \overline{\tau_L^n} = \overline{\frac{\int_0^\infty \! dt \, t^n R_{q,t}
    }{\int_0^\infty \! dt\, R_{q,t}}} \sim \int\! dU_L p(U_L) e^{nU_L/T},
\end{equation}
with $q=L^{-1}$ and the response function $R_{q,t-t'} =
\partial\Phi_q(t)/\partial\zeta_q(t')$.  The moment
$\overline{\tau_L^n}$ may behave as $\overline{\tau_L^n} \sim
e^{\alpha(n) U_0 L^\theta/T}$ with $\alpha(n)\geq n$, or grow even
faster with $\ln\overline{\tau^n} \gg L^\theta$.  A theory
of these timescales is crucial to understanding both equilibrium
response and correlations and to near-equilibrium phenomena such as
creep\cite{vortices,creepexp}.

Despite the wide applicability and extensive theoretical studies of
the elastic model, Eqs.~(\ref{eq:ham}-\ref{eom}), analytic treatment
for $d\geq 1$ of the non-zero temperature glassy behavior has been
limited to mean-field ($N\rightarrow\infty$) \cite{mezpar,frgN} and infinite dimension
($d\rightarrow\infty$) limits \cite{infinited} in which thermal activation over
divergent barriers ($U_L \sim L^\theta$) is not included.  The
existence of these growing barriers appears to be captured by the
field-theoretic Functional Renormalization Group (FRG) method 
\cite{fisher_functional_rg,frg_dep,frg2loop} (for
$\epsilon=4-d \ll 1$), extended to non zero temperature 
\cite{balents_loc,chauve_creep}, but until now neither the rare events nor
fluctuating barriers have been obtained in this approach.  

In this paper, we investigate how properties reminiscent of droplets
can emerge naturally from the FRG.  Although temperature is formally
irrelevant (in the RG sense), thermal effects give rise to an
intricate non-perturbative {\sl boundary layer} structure.
Nevertheless, a consistent scaling of this boundary layer
can be understood, providing a field-theoretic framework for
thermally-activated glassy dynamics.  This structure appears
to support solutions of the droplet type, with in particular rare
thermal fluctuations obeying Eq.~(\ref{eq:twowell}) and a distribution
of timescales $\overline{\tau^n}$ described by a non-trivial function
$\alpha(n)$.  Details can be found in \cite{us_long}.

We begin by considering the statics.  The theory is formulated in
terms of the disorder-averaged partition function for $n$ replicas
$\vec{\Phi}=(\Phi_1,\cdots,\Phi_n)$,
\begin{equation}
  Z_s = \!\!\int \!\!\!D\vec{\Phi} \exp\left\{ -\!\int_r \left[ \sum_a
      { \frac{1}{2T}} |\nabla \Phi_a|^2 \!-\! U[\vec{\Phi}(r)]
    \right]\right\}. \label{eq:staticPF} 
\end{equation}
A more general distribution for $V(\Phi,r)$ (still independently
distributed at each $r$) is expressed through the {\sl
  characteristic function} 
$U[\vec{\Phi}] = \sum_{k=2}^\infty {1 \over T^k k!} \sum_{a_1
    \cdots a_k} 
    S^{(k)}[\Phi_{a_1},\cdots,\Phi_{a_k}].$
Here
$S^{(k)}(\Phi_1,\cdots,\Phi_k)$ is the $k^{\rm th}$ connected cumulant
of $V(\Phi,r)$ at fixed $r$ (note
$S^{(2)}(\Phi,\Phi')=R(\Phi-\Phi')$).  By translational invariance,
these cumulants satisfy $S^{(k)}(\Phi_1+\lambda,\cdots,\Phi_k+\lambda)
= S^{(k)}(\Phi_1,\cdots,\Phi_k)$, so $U[\vec{\Phi}]$ is a fully
symmetric, translationally-invariant function of $\vec{\Phi}$. As is
well-known \cite{fisher_functional_rg}, {\sl zero-temperature}
power-counting shows that all $S^{(k)}$ for $k>2$ are
irrelevant, and the second cumulant $R(\Phi-\Phi')$ is marginally
relevant just below $d=4$.  One therefore finds that for $\epsilon=4-d
\ll 1$, the fixed point function $R^*(\Phi) \sim
\epsilon \ll 1$ and $\zeta$ also of $O(\epsilon)$.  This smallness
justifies the truncation $S^{(k)}\approx 0$ for $k>2$ at one-loop, to
leading order in $\epsilon$ at $T=0$.

Prior investigations of thermal effects have presumed this truncation
remains valid at $T>0$ \cite{balents_loc,chauve_creep}.  We check this
assumption by considering the corrections due to $S^{(3)}\neq
0$.  A straightforward Wilson scheme calculation
keeping only $k=2,3$ terms to one loop gives
\begin{widetext}
\vspace{-0.3in}
\begin{eqnarray}
  && \partial_l \tilde{R}(\Phi) =
  (\epsilon-4\zeta+\zeta\Phi\partial_\Phi)\tilde{R}(\Phi) +
  \overline{T}_l \tilde{R}''(\Phi) + \tilde{S}^{(3)}_{110}(0,0,\Phi) + 
  \frac{1}{2} \tilde{R}''(\Phi)^2 - \tilde{R}''(0) \tilde{R}''(\Phi)
  \label{BL2cum} 
  \\
  && \partial_l \tilde{S}^{(3)}(\Phi_{123}) = (2
  \epsilon-2-6\zeta+\zeta\Phi_i\partial_{\Phi_i}) 
  \tilde{S}^{(3)}(\Phi_{123}) + 
   \,{\rm
    sym}\left(\frac{3\overline{T}_l}{2}\tilde{S}^{(3)}_{200}(\Phi_{123})  
    + \frac{3\overline{T}_l}{2}\tilde{\sf R}''(\Phi_{13})
    \tilde{\sf R}''(\Phi_{23}) \right)     \label{BL3cum} \\  
  && + 3 \, {\rm sym}\left( \tilde{\sf R}''(\Phi_{12}) (
    \tilde{S}^{(3)}_{110}(\Phi_{113}) - 
    \tilde{S}^{(3)}_{110}(\Phi_{123})) 
    +\tilde{\sf R}''(\Phi_{12})\tilde{\sf R}''(\Phi_{13})^2  -
    \frac{1}{3}\tilde{\sf R}''(\Phi_{12}) \tilde{\sf R}''(\Phi_{23})
  \tilde{\sf R}''(\Phi_{31}) \right), \nonumber
\end{eqnarray}
\end{widetext}
\vspace{-0.2in}
\noindent where $\Lambda_l = \Lambda e^{-l}$ is the running 
cutoff, $\Phi_{ij}=\Phi_i-\Phi_j$,
$\Phi_{ijk}=\Phi_i,\Phi_j,\Phi_k$. With $A_d = 1/(2^{d-1} \pi^{d/2}
\Gamma(d/2))$, the rescaled functions 
$\tilde{R}(\Phi)=A_d \Lambda_l^{d-4+4\zeta} R(\Phi 
\Lambda_l^{-\zeta})$ and
$\tilde{S}^{(3)}(\Phi_{123}) = A_d^2 \Lambda_l^{2d-6+6\zeta}
S^{(3)}(\Phi_{123}\Lambda_l^{-\zeta})$.  The
ratio of thermal energy $T$ to pinning energy $L^\theta$  defines $\overline{T}_l = A_d T
\Lambda_l^\theta$, 
where $\theta=d-2+2\zeta>0$ indicates the dominance of energy over entropy as
appropriate for a glass.  It was convenient to define $\tilde{\sf
  R}(\Phi) = \tilde{R}(\Phi)-\tilde{R}''(0) \Phi^2/2$, since
$\tilde{R}''(0)$ does not feed into higher cumulants.

The naive $T>0$ analysis assumes $\tilde{S}^{(3)} \ll
\epsilon\tilde{R}$ to convert Eq.~(\ref{BL2cum})\ into a decoupled
equation for $\tilde{R}(\Phi)$.  Its solution yields two regimes.  For
$\Phi \sim O(1) \gg \overline{T}_l$, the solution $\tilde{R}_l(\Phi)$
converges to a zero temperature fixed point function
\cite{fisher_functional_rg}, with a small set of universality
classes.\cite{vortices,balents_loc} For, e.g. $V(\Phi,x)$ periodic on
the interval 
$0<\Phi<1$, the force-force correlator 
$\tilde\Delta(\Phi)=-\tilde{R}''(\Phi)={\rm Min}_{n\in {\cal Z}}
{\epsilon\over 6}[(\Phi-n-1/2)^2-1/12]$.  {\sl Non-analytic} behavior
at small $\Phi$, $\tilde\Delta_{T=0}(\Phi)-\tilde\Delta(0) \sim -\chi
|\Phi|$ ($\chi\equiv|\tilde\Delta'(0^+)|$) is, however, {\sl
  super-universal}, i.e.  the same for all disordered elastic models.
The significance of this cusp has been discussed by several
authors.\cite{fisher_functional_rg,balents_rsb_frg,frgN}\ Notably, the
fluctuations of the mean curvature of the pinning potential,
$-\tilde\Delta''_{T=0}(0)=\overline{(\partial_\Phi^2
  V(\Phi))^2}=+\infty$, which implies the existence of multiple
metastable minima in the effective potential.  We will see further
consequences below.  In this $T=0$
regime, the neglect of $S^{(3)}$ appears consistent: solution of
Eq.~(\ref{BL3cum}) gives $\tilde{S}_{T=0}^{(3)} \sim O(\epsilon^3) \ll
\tilde{R}_{T=0}$, yielding only higher-order corrections.

The convergence to this solution is non-uniform, however, and for
$\Phi \alt \epsilon\overline{T}_l$, the cusp singularity is rounded
in a {\sl thermal boundary-layer}.  The boundary layer form follows
from Eq.~(\ref{BL2cum})\ for $\hat{S}^{(3)}=0$ through the 
ansatz $\tilde{\sf R}(\Phi)= - \overline{T}_l^3 (\epsilon\tilde\chi)^{-2}
r(\phi)$, with $\tilde{R}''(0)=\tilde{\chi}^2
\epsilon^2/(2\zeta-\epsilon)$, keeping $\phi=\epsilon\tilde{\chi}
\Phi/\overline{T}_l$ fixed of $O(1)$, $\tilde{\chi}=\chi/\epsilon$ and
$\overline{T}_l \ll \epsilon^2$.  This gives
$r^{\prime\prime}(\phi)=1-\sqrt{1+\phi^2}$,\cite{chauve_creep}\ matching
the $|\phi|$ cusp as $\phi\rightarrow \infty$, but becoming
smooth for $\phi$ of $O(1)$.

Even within the one loop Wilson RG, however, the neglect of the third
cumulant is unjustified in the boundary layer.  This follows since the 
$O[({\sf R}'')^3]$ terms in Eq.~(\ref{BL3cum}) feed their boundary-layer
scaling forms into $S^{(3)}$.  Naively balancing these terms with the
rescaling part ($\approx -2 \tilde{S}^{(3)}$) suggests
$\tilde{S}^{(3)} \sim \overline{T}_l^3$.  This is, however, inconsistent,
since the remaining terms linear in $\tilde{S}^{(3)}$ would then be
$O(\overline{T}_l^2)$.  Instead,  the only consistent ansatz is
$\tilde{S}^{(3)}(\Phi_1,\Phi_2,\Phi_3) =
(\epsilon\tilde\chi)^{-2}\overline{T}_l^4 
s^{(3)}(\phi_1,\phi_2,\phi_3)$.  This in turn generates a feedback in
Eq.~(\ref{BL2cum}) that is
of the same order as the other surviving terms in the boundary layer.

In fact, the scaling forms in
Eqs.~(\ref{BL2cum}-\ref{BL3cum}) must be extended to {\sl all}
cumulants.  Defining $U(\vec\Phi) = A_d \Lambda_l^d
\tilde{U}(\vec{\Phi} \Lambda_l^\zeta)$, then
$\tilde{U}(\vec{\Phi}) = \sum_{p, \{ a_i \}}
  f_{2p}
  \phi_{a_1}\cdots\phi_{a_{2p}} +
  \frac{\overline{T}_l}{(\epsilon\tilde\chi)^2}  u( \vec{\phi} )$, 
with  $f_{2p} =
\tilde{\Delta}^{(2p)}(0)/[(2p)!(\tilde\chi\epsilon)^{2p}]$, where 
$\tilde{\Delta}^{(2p)}(0)$ is the $(2p)^{\rm th}$ cumulant of the random force
($\tilde{\Delta}^{(2)}(0)=\Delta(0)$), and
the boundary-layer replica vector $\vec{\phi}_l =
{\epsilon\tilde{\chi} \vec{\Phi}}/\overline{T}_l$.  This gives the
non-trivial one-loop Wilson fixed point equation,
\begin{equation}
  \!\!\sum_{p, \{ a_i \}}\!\!
  x_{2p} f_{2p}
  \phi_{a_1}\!\cdots\phi_{a_{2p}}  \!\!+ {\rm Tr} \!
  \ln \left[ \delta^{ab} -  \partial_a \partial_b 
    u(\vec{\phi}) \right] = 0, \label{eq:BL2}
\end{equation}
with scaling eigenvalues $x_2=\epsilon-2\zeta$ of $O(\epsilon)$
and $x_{2p} = 
d-2p(\theta-\zeta)$ of $O(1)$ for $p>1$.
Expansion of Eq.~(\ref{eq:BL2}) in number of replica sums gives
coupled boundary layer equations for $r$ and $s^{(3)}$, and all higher
cumulants. 

Unfortunately, the disappearance of $\epsilon$
from Eq.~(\ref{eq:BL2}) (since $x_{2p} f_{2p} \sim O(1)$),
indicates that a one-loop
analysis may be insufficient.  Worse, higher loop corrections to $U$ appear naively extremely singular in $1/T$:
each additional contraction carries with it two $\partial_\Phi$
derivatives, and hence two factors of $1/T$, but only a single
propagator proportional to $T$, so increasing powers of $1/T$ appear
in higher loop diagrams.  This makes the non-perturbative status of
the boundary layer extremely nontrivial.  To verify its existence, we
therefore employ the {\sl Exact RG} (ERG) method \cite{erg}, taking
from now on $\zeta=0$ for simplicity.  The ERG
extends the characteristic function $U(\vec\Phi)$ to a non-local {\sl
  functional}, ${\cal U}[\vec{\Phi}(r)]$, defined by a
running effective action $\Gamma(\vec{\Phi}) = (1/2) \Phi : G_l^{-1} :
\Phi + {\cal U}_l[\vec{\Phi}]$, with a cutoff Green's function
$(G_l)_{ab}(q)=\delta_{ab}(T/q^2) (c(\frac{q^2}{\Lambda^2}) - c(\frac{q^2}{\Lambda_l^2}))$.  Then
\begin{equation}
  \label{eq:ERG}
  \partial_l {\cal U}_l(\vec{\Phi}) = \frac{1}{2} {\rm Tr}\,
  \partial_l G_l : (G_l^{-1}\!-\! G_l^{-1}(1\!+\!G_l: \frac{\delta^2 {\cal
      U}_l}{\delta \Phi \delta \Phi})^{-1}),
\end{equation}
where $\delta's$ indicate functional differentiation.  This equation
can be expanded order by order in ${\cal U}_l$ to generate a series
expansion.  The functional nature of ${\cal U}[\vec\Phi(r)]$
necessitates a further {\sl multilocal} expansion ${\cal
  U}_l[\vec\Phi]=\sum_{p=1}^\infty \int_{x_1\cdots x_p}
U_l^{(p)}(\vec\Phi_{x_1}\cdots\vec\Phi_{x_p};x_1\cdots x_p)$, in terms
of translationally-invariant {\sl functions} (``p-local terms'') of
$p$ $\Phi_i$ and $x_i$ variables.  Remarkably, {\sl exact}
integro-differential flow equations for the $U_l^{(p)}$ can be
obtained.  The boundary layer ansatz, applied to each p-local term
generalizes to $U_l^{(p)}(\{ \vec\Phi_i\},\{x_i\}) =
\epsilon^{-2}\Lambda_l^{p d} \overline{T}_l u^{(p)}_l(\{
\vec\phi_{il}\} , \{ \Lambda_l x_i \} )$, plus a nonlocal extension of
the random-force terms ($f_{2p}$).  Inserting this into
Eq.~(\ref{eq:ERG}), one finds that apart from the random-force terms,
the left-hand-side is negligible for $\overline{T}_l \ll 1$ and
$\vec\phi\sim O(1)$.  This gives a non-trivial hierarchy of
integro-differential equations (too complicated to present here) for
the scaling functions $u_l^{(p)}$ {\sl with a well-defined limit as $l
  \rightarrow \infty$}.  We have checked that the truncation to $p\leq
2$ is non-singular \cite{us_long}, i.e. the $p=2$ term can be
eliminated to obtain a single equation for the local function
$u_l^{(1)}$ involving no divergent integrals. We believe this
procedure, which effectively resums the $1/T$ divergences into a
non-singular result, can be carried out systematically to all orders. 

This indicates the non-perturbative consistency of the boundary layer
ansatz in the ERG.  We now turn to some physical implications.
A tree level calculation gives
\begin{equation}
    \overline{(\langle \Phi^2\rangle - \langle \Phi\rangle^2)^n} \sim
    T^{2n} \partial^{2n}_{\{\Phi_a\}}U(0) \int d^d{\bf r}'\, G^{2n}({\bf r'}),
\end{equation}
where $\partial^{2n}_{\{\Phi_a\}}U$ is a combination of 
$2n^{\rm th}$ order derivatives of $U(\vec{\Phi})$, and $G({\bf r'})=
\int d^d{\bf q}/q^2 \sim |r'|^{2-d}$.  Remarkably, inserting the boundary layer
scaling form precisely reproduces the droplet behavior of thermal
fluctuations, Eq.~(\ref{eq:twowell}), with $c_n
\sim \partial^{2n}_{\{\phi_a\}}u(0)$! A further test of droplet predictions
is that the simplest {\sl three-well} quantity $C_3=\langle
(\Phi_a-\Phi_b)^2(\Phi_b-\Phi_c)^2(\Phi_a-\Phi_c)^2\rangle$ vanishes
to $O(T)$. We have verified $C_3 \sim
O(T^2)$ for $d=0,N=1,\theta>0$. Some further 
results\cite{us_long} agree with
recent calculations in this case
where the simplest droplet picture is exact \cite{toy}.

We now turn to the dynamics.  Prior
investigations \cite{chauve_creep}\ have calculated 
the renormalization of the {\sl mean} relaxation time at
$O(\tilde\Delta)$, $\partial_l \overline{\eta}_l =
-\tilde\Delta''(0)\overline{\eta}_l$, leading through the boundary
layer scaling $\tilde{\Delta}''(0) \sim 1/\overline{T}_l$ to $\overline\eta_l
=\eta_0 \exp[ \frac{\chi^2}{\theta} (1/\overline{T}_l - 1/\overline{T}_0)]$ in Eq.~(\ref{eom}).
This is consistent with an energy barrier $U_L = U_{0} L^{\theta}$
(with $L=\Lambda_l^{-1}$).

We begin our consideration of the {\sl distribution} of relaxation
times in a simplified calculation
generalizing Eq.~(\ref{eom}) to a random
friction coefficient $\eta\rightarrow \eta(r)$, where the
fluctuation-dissipation theorem requires $\langle \zeta(r,t)
\zeta(r',t') \rangle = 2 \eta(r) T \delta^d(r-r')\delta(t-t')$.  We
take $\eta(r)$ with an independent and identical distribution
$P(\eta)$ at each $r$, with the characteristic function
$F(z)=-\ln\int_0^{+\infty} d \eta P(\eta) e^{- z \eta}$.  In the
standard Martin-Siggia-Rose (MSR) 
formalism, thermal and disorder averages over the
equation of motion are calculated from a functional integral over
$\Phi_{rt},\hat\Phi_{rt}$ with the measure $\exp(-S_0-S_{\rm
  st}-S_{\rm dy})$, where $S_0= \int_{r t} - i \hat{\Phi}_{rt}
\nabla^2_{r} \Phi_{rt}$,
\begin{eqnarray}
  \label{smod1} 
 S_{\rm st} &  \!\! = \!\! &  \sum_n \!\frac{-1}{n!} \!\!\int_{rt_1\cdots t_n}
  \!\!\!\!\!\!\!\!\!\!\!\!\!(i
  \hat{\Phi}_{rt_1}\!\!\cdots i
  \hat{\Phi}_{rt_n})\Delta^{(n)}(\Phi_{rt_1},\cdots,\Phi_{rt_n}\!),
    \\
  S_{\rm dy}  &  \!\! = \!\! &  \int_{r}
  F[ \int_t
  ( i \hat{\Phi}_{rt} \partial_{t} \Phi_{rt} - T i \hat{\Phi}_{rt} i
  \hat{\Phi}_{rt})],  \label{smod2}
\end{eqnarray}
and $\Delta^{(n)}(\cdot) = (-1)^n
\partial_{\Phi_1}\cdots\partial_{\Phi_n} U^{(n)}(\cdot)$.  Setting
$F(z)=\overline{\eta}z$ recovers the previous approach, but when
$F(z)$ is non-trivial Eqs.~(\ref{smod1}-\ref{smod2}) comprise an
unusual dynamical field theory.  First consider the {\sl random
  friction model} with $\Delta^{(n)}=U^{(n)}=0$. Despite its
highly non-quadratic action, this represents an FRG fixed manifold,
with non-trivial response function  
moments, indexed by the ``F-term'' $F(z)$, which remains
un-renormalized.  This follows since Eq.~(\ref{eom})
becomes linear for zero pinning force $f=0$, hence the distribution of
$\eta(r)$ is scale-independent.

This changes drastically in the presence of pinning disorder
$\Delta^{(n)}\neq 0$.  Considering for the moment only the second
cumulant $n=2$, to $O(\Delta)$ $F(z)$ is no longer scale-independent:
$\partial_l F_l[z] = (\partial_l \ln \overline{\eta}_l) ( z F_l'[z] +
2 z^2 F_l''[z])$.  This implies the rapid growth of higher connected
cumulants of $\eta$, to wit $F(z)=-\sum_q \frac{(-1)^q}{q!} \eta^{(q)}
z^q$, with 
\begin{equation}
  \eta^{(q)}_l \sim \eta^{(q)}_0
  e^{\chi^2 \alpha(q)/\overline{T}_l \theta} = 
\eta^{(q)}_0 (\overline{\eta}_l/\eta_0)^{\alpha(q)}
\gg
  (\overline{\eta}_l)^q , \label{Arrhenius}
\end{equation}
where $\alpha(q)= 2 q^2 - q$.\cite{us_long,FR}\ Thus all the
renormalized moments of the friction have the Arrhenius form, but
since $\alpha(q) > q$, they cannot be characterized by a single
relaxation time $\overline{\eta}_l$.  The quadratic form of
$\alpha(q)$ obtained in this approximation is consistent with a very
broad, log-normal distribution of relaxation times $\eta$.  The more
complete analysis below provides a consistent ansatz which extends
this scaling to the full effective action, with, however, non-trivial
exponents $\alpha(q)$.\cite{zerodnote}  

To go beyond this simplified picture, we must consider all interaction
vertices at a given ``level'' $q$ involving a fixed number $q$ of
total time derivatives (power of frequency $\omega^I$), which includes
e.g.  $F^{(q)}[0]$ but also an infinite hierarchy of higher cumulant
functions, as in the static case $q=0$.  An examination of the
one-loop low-temperature FRG shows that for each value of
$q$ one obtains a distinct hierarchy such that all terms grow
proportionately to $\eta_l^{(q)}$ with a non-trivial exponent
$\alpha(q)$.  The hierarchies at different $q$ are coupled in a
special manner such that only levels $q'<q$ feed into level $q$.  This
structure allows for {\sl independent} exponents $\alpha(q)$ provided
they satisfy the natural convexity constraint $\alpha(q) >
\alpha(q')+\alpha(q-q')$ for $1\leq q' \leq q-1$.  Each member of
the hierarchy scales as $\eta_l^{(q)}$ times a distinct
scaling function in the boundary layer regime $\phi\sim O(1)$.
Coupled differential eigenvalue equations involving these functions
determine the exponents $\alpha(q)$ by matching.

We now illustrate this structure for $q=1$, calculating the FRG for
terms up to second cumulant order: 
\begin{eqnarray}
  \label{eq:dynints}
  \!\!\! S_{\rm dy} & \!\!=\!\! &  \int_{r t} \!\!\overline{\eta} \,i
  \hat{\Phi}_{rt}  
      \dot{\Phi}_{rt} \!-\!\! \int_{r t_1
      t_2} \!\!\!\!\!\!\!\!\!\!
    i\hat{\Phi}_{rt_1} i\hat{\Phi}_{rt_2} \dot{\Phi}_{rt_1}
    G(\Phi_{rt_1}\!\!\!-\!\Phi_{rt_2}) \nonumber \\
    & & \hspace{-0.2in} -\frac{1}{6} \int_{rt_1t_2t_3} \!\!\!\!\!\!
    i\hat{\Phi}_{rt_1} i\hat{\Phi}_{rt_2} 
    i\hat{\Phi}_{rt_3} \dot{\Phi}_{rt_1}
    H(\Phi_{rt_1},\Phi_{rt_2},\Phi_{rt_3}) 
  \end{eqnarray}
where $G(-\Phi)=-G(\Phi)$.  One finds that $G$ 
contributes to the renormalization of $\overline\eta$:
\begin{eqnarray}
&&  \partial_l \overline{\eta} = ( \overline{G}'(0) -
\tilde{\Delta}''(0) ) \overline{\eta}, \label{eq:etabar}
\end{eqnarray}
where
$G(\Phi)=\overline{\eta}_l\Lambda_l^{2-d}A_d^{-1}\overline{G}(\Phi)$,
correcting the naive 
$\tilde{\Delta}''(0)$ contribution to
$\overline{\eta}$.  Although $\overline{G}$ is nominally irrelevant,
$\partial_l \overline{G}  = (-2 + \epsilon) \overline{G}+\cdots$,
it supports a non-trivial boundary
layer, in which $\overline{G}(\Phi) = \epsilon \tilde\chi g(\phi)$, satisfying
\begin{eqnarray}
 0 & \!\!=\!\!  & 2 f'' g + (1+f) g''  - g'(0) (g- f') 
 +  2f' (f''(0) + f'') \nonumber \\ 
 &  &  \hspace{-0.2in}+ 3 g' f'+ s^{(3)}_{221}(\phi,\phi,0) + 2
 h_{100}(0,\phi,0) + \frac{2}{3}h_{010}(0,\phi,0), \nonumber   \label{eq:BLdyn}
\end{eqnarray}
where $f(\phi)=-r''(\phi)$, $H(\Phi_{123}) = \overline{\eta}_l
\Lambda_l^{4-2d} A_d^{-1} \epsilon^2 h(\phi_{123})$.  Matching to the
well-behaved ``outer'' solution requires $g(\phi)
\rightarrow_{\phi\rightarrow \infty} g_{+\infty}$, a constant, while
for $\phi \ll 1$, $g(\phi) \sim g'(0)\phi$.  For given functions
$f,s^{(3)},h$, this equation is an eigenvalue problem for $g'(0)$
(indeed, analysis shows that $g'(0)$ must be tuned to match the outer
solution).  From Eq.~(\ref{eq:etabar}), the exponent
$\alpha(1)=g'(0)+f''(0)$ is thereby non-trivial as claimed.  Moreover,
$\alpha(1)$ is determined through the above equation in a manner
depending upon the full boundary layer functions $r$, $s$, $h$.
Similar considerations apply to $q=2$ (details suppressed to save
space), for which the one-time term is $D\int_{rt}\hat{u}\ddot{u}$,
and there are $3$ independent functions at the second cumulant level.
The $q=1$ functions feed into the $q=2$ hierarchy in such a way that
the $\alpha(2)$ exponent governing the growth of $D$ is independent
provided $\alpha(2)>2\alpha(1)$ self-consistently.

In summary, we have described a consistent non-perturbative boundary
layer structure in a static field theory for pinned elastic systems,
successfully describing rare events dominating low-temperature thermal
fluctuations.  An extension to the dynamics leads to a broad distribution of
relaxation times, for which the moments
$\overline{\tau^n}$ retain a (non-trivial) Arrhenius form.  This
condition would appear required for simple scaling arguments for creep
to be even qualitatively correct, and is experimentally testable.  Our
investigations uncover an intimate connection between the
highly non-perturbative boundary layer and droplet
considerations, suggesting that an analytic approach encapsulating
this physics might be successful in the near future. The more general
applicability of these ideas to more complex glassy systems such as
spin and structural glasses, as well as to the quantum regime, appears
as a distinct possibility.

L.B. thanks L. Radzihovsky and D. S. Fisher for inspiration.  L.B. was
supported by NSF grants DMR-9985255,INT-0089835, and the Sloan and
Packard foundations.



\begin{thebibliography}{999}
\bibitem{zerodnote} In $d=0$ for $N=1$
scaling of $\overline{\tau_L^n}$ is non-Arrhenius\cite{toy},
indicating a different scenario there.
\end{thebibliography}
\end{document}